%% file: MBndm.tex
\documentclass[aps,prl,twocolumn,superscriptaddress,showpacs,amsmath]{revtex4-1}
\usepackage{graphicx}  
\usepackage[caption=false]{subfig}
\graphicspath{{figs//}}
\DeclareGraphicsExtensions{.pdf,.png}
\usepackage{bm}            
\usepackage{amssymb}   
\usepackage{amsmath}   

\def\nubar     {\overline{\nu}}
\def\epalp     {\epsilon^4\alpha_D}

\begin{document}

\title{
Dark Matter Search in a Proton Beam Dump with MiniBooNE}
\input MBndmAuthors.tex
\date{\today}

\begin{abstract}
The MiniBooNE-DM collaboration searched for vector-boson mediated production of dark matter using the Fermilab 8 GeV Booster proton beam in a dedicated run with $1.86 \times 10^{20}$ protons delivered to a steel beam dump.  The MiniBooNE detector, 490~m downstream, is sensitive to dark matter via elastic scattering with nucleons in the detector mineral oil. Analysis methods developed for previous MiniBooNE scattering results were employed, and several constraining data sets were simultaneously analyzed to minimize systematic errors from neutrino flux and interaction rates. No excess of events over background was observed, leading to a 90\% confidence limit on the dark-matter cross section parameter, $Y=\epsilon^2\alpha_D(m_\chi/m_V)^4 \lesssim10^{-8}$, for $\alpha_D=0.5$ and for dark-matter masses of $0.01<m_\chi<0.3~\mathrm{GeV}$ in a vector portal model of dark matter. This is the best limit from a dedicated proton beam dump search in this mass and coupling range and extends below the mass range of direct dark matter searches.  These results demonstrate a novel and powerful approach to dark matter searches with beam dump experiments.
\end{abstract}

\maketitle

\noindent \textbf{Introduction --- }
There is strong evidence for dark matter (DM) from observations of gravitational phenomena across a wide range of distance scales~\cite{Olive:2016xmw}. A substantial program of experiments has evolved over the last several decades to search for non-gravitational interactions of DM, with yet no undisputed evidence in this sector. 
Most of these experiments target DM with weak scale masses and are less sensitive to DM with masses below a few GeV. To complement these approaches, new search strategies sensitive to DM with smaller masses should be considered~\cite{Alexander:2016aln}.

\begin{figure}
\includegraphics[width=0.5\textwidth]{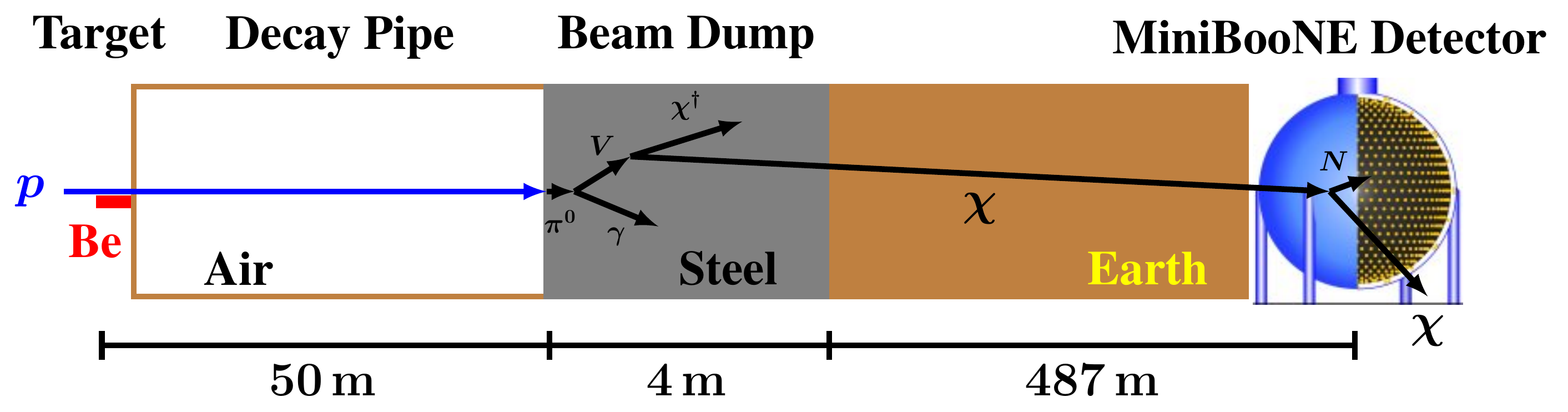}
\caption{\label{fig:beamOffTarget} Schematic illustration of this DM search using the the Fermilab BNB in off-target mode together with the MiniBooNE detector. The proton beam is steered above the beryllium target in off-target mode lowering the neutrino flux.}
\end{figure}

Fixed-target experiments using beams of protons or electrons can expand the sensitivity to sub-GeV DM that couples to ordinary matter via a light mediator particle~\cite{Batell:2009di,deNiverville:2011it,deNiverville:2012ij,Dharmapalan:2012xp,Izaguirre:2013uxa,Izaguirre:2014bca,Izaguirre:2014dua,Batell:2014yra,Batell:2014mga,Dobrescu:2014ita,Soper:2014ska,Kahn:2014sra,Izaguirre:2015yja,deNiverville:2015mwa,Izaguirre:2015pva,Coloma:2015pih}. In these experiments, DM particles may be produced in collisions with nuclei in the fixed target, often a beam dump, and may be identified through interactions with nuclei in a downstream detector.  Results from past beam dump experiments have been reanalyzed to place limits on the parameters within this class of models. In this Letter, we report on the first dedicated search of this type (proposed in~\cite{Dharmapalan:2012xp}), which employs 8~GeV protons from the Fermilab Booster Neutrino Beam (BNB), reconfigured to reduce neutrino-induced backgrounds, combined with the downstream MiniBooNE (MB) neutrino detector  (Fig.~\ref{fig:beamOffTarget}).

A DM particle may couple to ordinary matter through a light mediator particle which could also control interactions with Standard Model particles, allowing the correct relic abundance in the standard thermal freeze-out scenario~\cite{Batell:2009di,deNiverville:2011it,deNiverville:2012ij}. A minimal dark sector model of this type is known as {\em vector portal DM}~\cite{Pospelov:2007mp,ArkaniHamed:2008qn} and is used as a framework for the analysis presented here. Although we emphasize that this search is sensitive to other scenarios, in this particular one, interactions of $\chi$ are mediated by a $U(1)$ gauge boson $V_\mu$  (``dark photon'') that kinetically mixes with the ordinary photon. Four unknown parameters control the physics: DM mass $m_\chi$, $V_\mu$ mass $m_V$, kinetic mixing $\epsilon$, and dark gauge coupling $g_D$. For this work, the DM particle is assumed to be a complex scalar, which is consistent with terrestrial, astrophysical, and cosmological constraints~\cite{deNiverville:2012ij}.

\begin{figure}
  \centering
 \subfloat[$\pi^0$, $\eta$ Decay]{\includegraphics[width=0.22\textwidth]{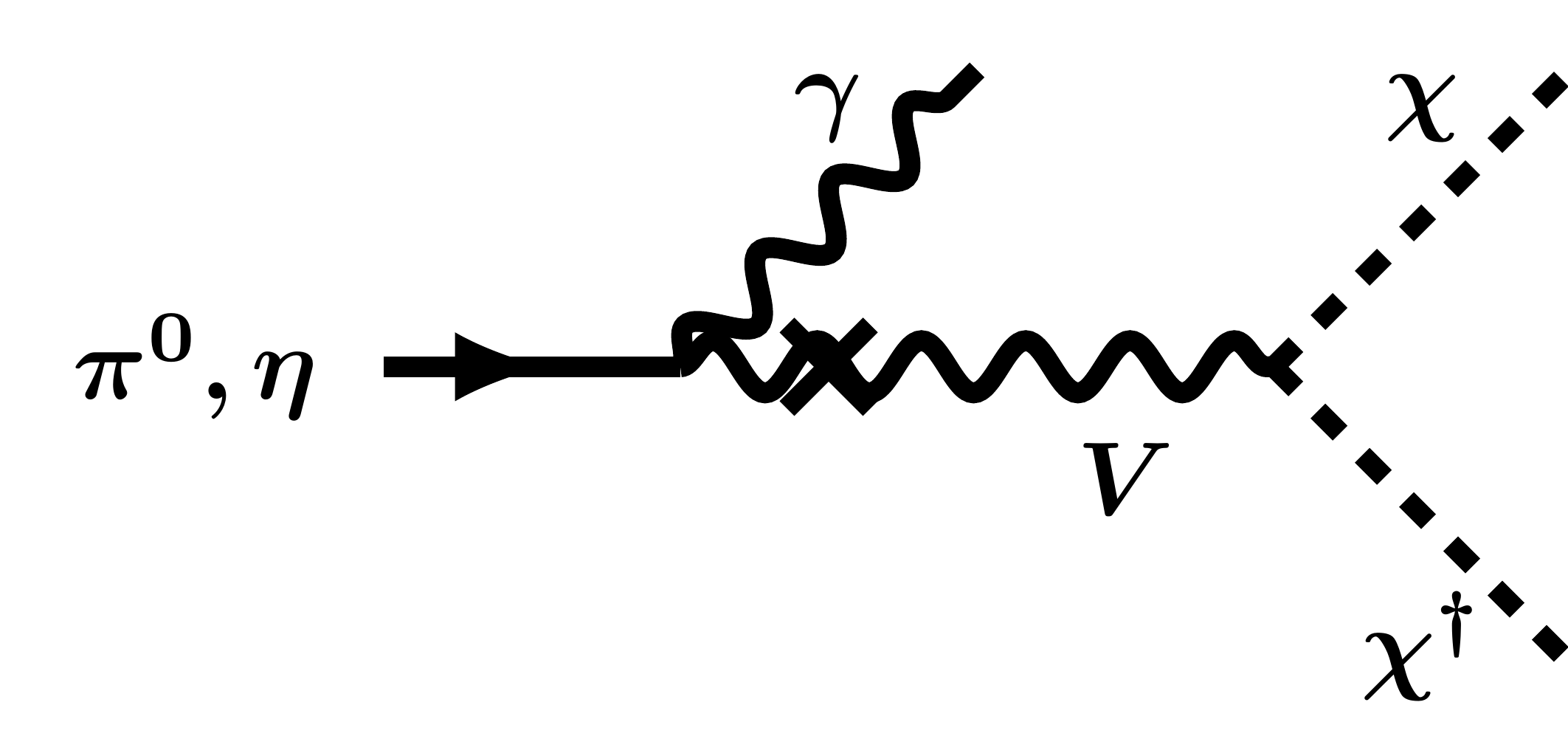}}
\quad
 \subfloat[Proton Bremsstrahlung]{\includegraphics[width=0.22\textwidth]{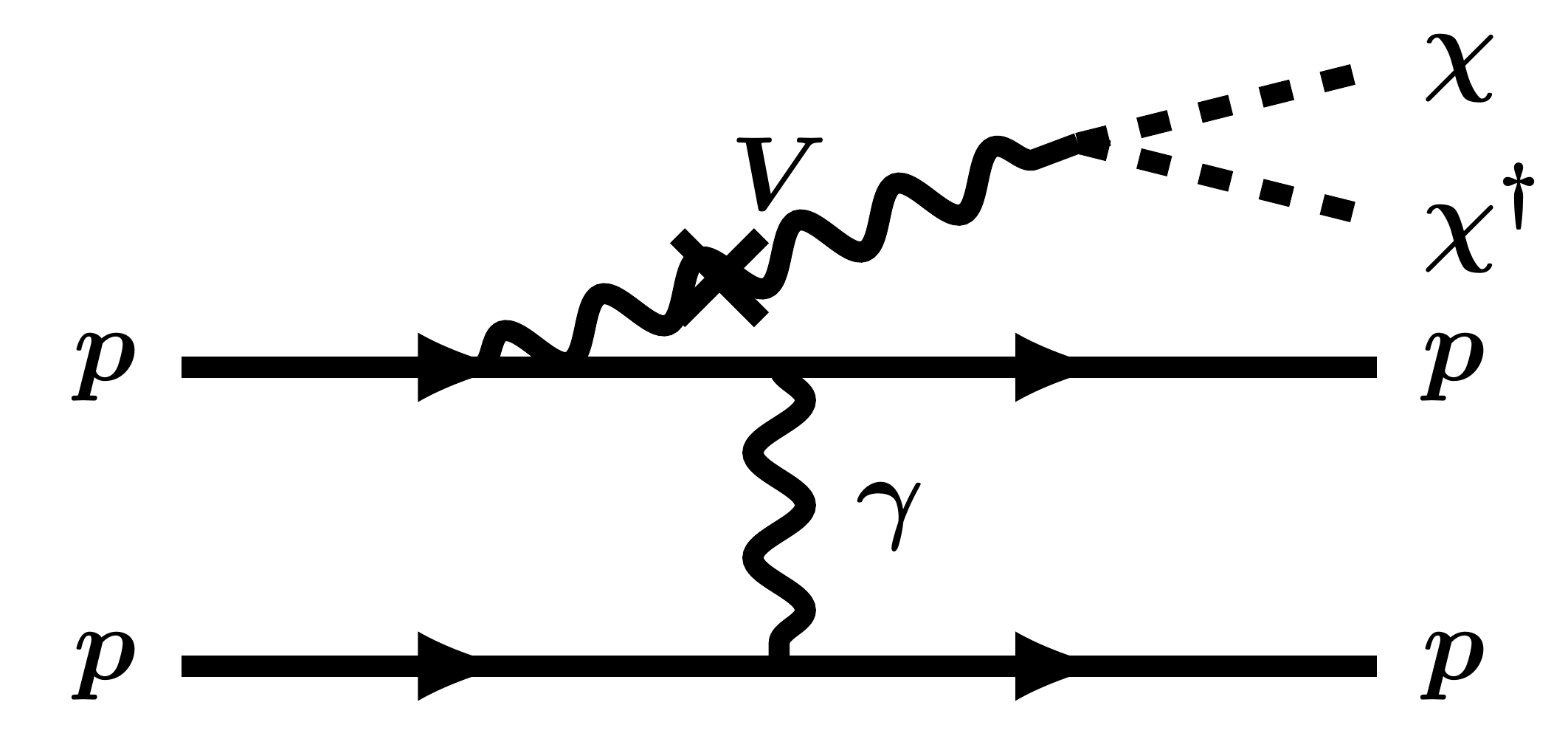}}
  \caption{\label{fig:theoryFeynProd} DM production channels relevant for this search with an 8~GeV proton beam incident on a steel target.}
\end{figure}

\begin{figure}
  \centering
  \subfloat[Free protons]{\includegraphics[width=0.22\textwidth]{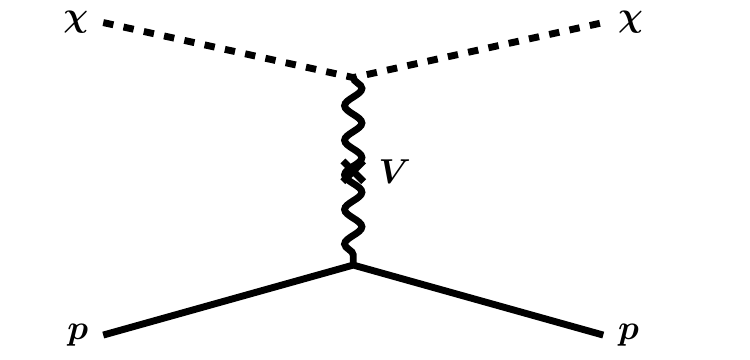}}
   \quad
   \subfloat[Bound nucleons in $^{12}C$.]{\includegraphics[width=0.22\textwidth]{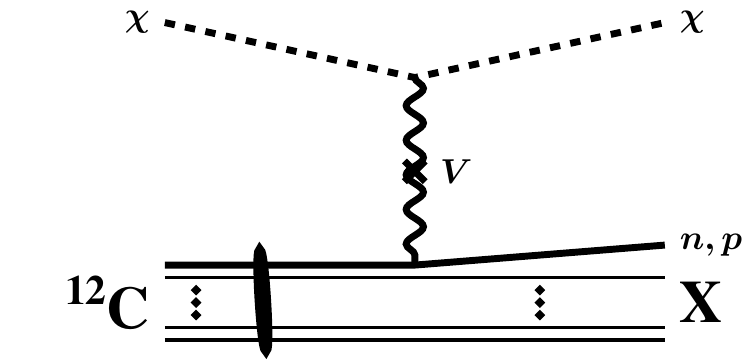}}
  \caption{\label{fig:theoryFeynInter} DM interactions with nucleons in the detector.}
\end{figure}

Two different DM production mechanisms (Fig.~\ref{fig:theoryFeynProd}) likely dominate for this search: 1) decay of secondary $\pi^0$ or $\eta$ mesons and 2) proton bremsstrahlung.  For both of these processes, the production rate scales as $\epsilon^2$ provided the $V_\mu$  can decay into two on-shell DM particles with $m_V > 2 m_\chi$.  The $\chi$, produced via one of these mechanisms, may be detected via interactions with nucleons or electrons. This search is sensitive to DM-nucleon interactions $\chi N$, mediated by $V_\mu$ exchange (Fig.~\ref{fig:theoryFeynInter}) and the scattering rate in the detector scales as $\epsilon^2 \alpha_D$, where $\alpha_D = g_D^2/4\pi$. Combining this with the production rate behavior yields a DM event rate that scales as $\epsilon^4\alpha_D$ for $m_V > 2 m_\chi$. 

\noindent \textbf{Experiment ---}
In the neutrino-production mode (``$\nu$-mode'') configuration of the BNB, 8~GeV protons from the Fermilab Booster are delivered to a 1.75-interaction-length beryllium target  in pulses with intensity $3-5\times10^{12}$ protons and 1.6~$\mu$s in duration, creating a large flux of charged mesons, predominantly pions.  A magnetic horn surrounds the target and uses a pulsed $\approx$ 1.5~T magnetic field to guide the mesons down a 1~m radius, 50~m long cylindrical, air-filled, decay pipe that terminates into a steel beam stop. The majority of mesons decay into neutrinos (e.g. $\pi \rightarrow \mu\nu$) providing a large neutrino flux in the downstream detector~\cite{AguilarArevalo:2008yp}. 

For this DM search, the beamline was configured in ``off-target'' mode with the 8~GeV protons steered off of the beryllium production target, through the powered-off magnetic horn, and into the steel beam dump at the end of the decay region.  This greatly reduces the flux of neutrinos created via meson decay in-flight, thus lowering the neutrino event background.  This increases sensitivity to DM produced in decays of $\pi^0$ and $\eta$, which are produced copiously in the beam dump. 

The flux of neutrinos and associated errors in $\nu$-mode were calculated using experimental data along with a simulation program detailed in \cite{AguilarArevalo:2008yp}.  To predict the off-target flux, the simulation was updated with the addition of various beam line components that are important only for off-target running. These additional components have negligible effects in $\nu$-mode as the beryllium target and surrounding aluminum is the source of 99\% of the mesons contributing to the neutrino flux at the detector.  However, in off-target mode, only $\approx 30\%$ of the mesons resulting in detector neutrinos are created in the beryllium target and surrounding aluminum, so other beam-line materials are important.  The beam parameters (direction, emittance, lateral size, etc.) used by the simulation were measured during the run. 

Charged-current quasielastic (CCQE) scattering of muon-neutrinos produces a readily detected muon and is the highest-rate neutrino process in the MB detector. With the assumption that DM scattering is purely elastic, the CCQE samples are free of DM-scattering events and, since they are well-measured via the large samples gathered in $\nu$-mode running, can be used to constrain the off-target neutrino flux.   A sample of 956 CCQE events from off-target mode were reconstructed and compared to that predicted by the beam and detector simulations. The beam parameters input to the simulation were then adjusted, within their uncertainties, to reproduce that number of events and to improve the off-target flux estimate.   A set of beam simulation variations, consistent with errors on the beam parameters and the total number of CCQE events, was created in order to determine the error on predicted fluxes. 

\begin{figure}
\centering
\includegraphics[width=0.99\columnwidth]{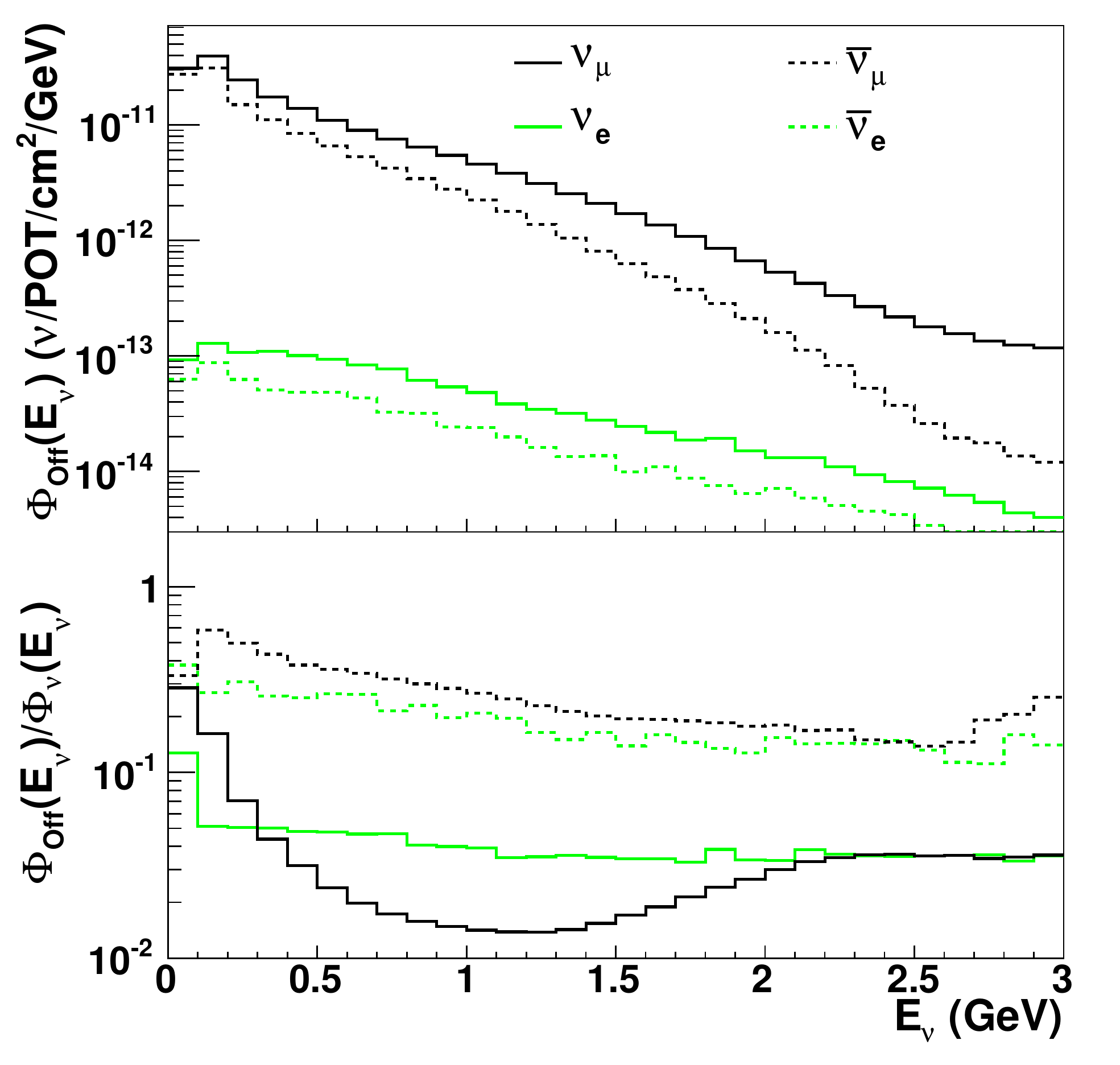}
\caption{\label{fig:flux} Predicted flux (top) in off-target mode and the flux ratio for off-target to $\nu$-mode (bottom) as a function of neutrino energy for each neutrino species.}
\end{figure}

The resulting predicted neutrino flux for off-target mode is shown in Fig.~\ref{fig:flux} along with the ratio of off-target flux to that for $\nu$-mode.  The predicted off-target flux for $0.2<E_\nu<3$~GeV is $(1.9\pm1.1)\times10^{-11}~\nu$ POT$^{-1}$ cm$^{-2}$ (``POT'' is proton-on-target).  The mean energy of the off-target neutrino flux is 660~MeV compared to 830~MeV in $\nu$-mode. The integrated off-target flux is 1/27 of the $\nu$-mode flux and the event rate 1/48 that of $\nu$-mode. The total data set reported here used $1.86\times10^{20}$ POT collected from Nov. 2013--Sept. 2014.

During this run period, the MB detector operated as for the previous 12 years which has included searches for neutrino oscillations and measurements of neutrino cross sections in both $\nu-$ and $\nubar-$mode.  In particular, MB has measured $\nu$ and  $\nubar$-nucleon neutral-current elastic (NCE) scattering~\cite{AguilarArevalo:2010cx, Aguilar-Arevalo:2013nkf} which has the same expected  final state as $\chi N$ scattering, allowing for the same mode of operation with well-understood detection and analysis methods. 

The MB detector~\cite{AguilarArevalo:2008qa}  consists of 818 tons of mineral oil (CH$_2$) in a 610~cm-radius tank viewed by 1280 photomultiplier tubes (PMTs) in the inner, primary region and 240 PMTs arranged in pairs viewing the outer, optically separate, 35~cm thick veto region.  Any PMTs  with signal $>0.1$ photoelectron  are digitized and recorded in a 19.2~$\mu$s time window around the 1.6~$\mu$s BNB proton pulses.  The signature of $\chi N$ scattering events is a pattern of hits consistent with a track from a single proton or neutron of a few hundred MeV kinetic energy. The MB detector is sensitive to these sub-Cherenkov particles via a small amount of scintillation light emitted as they traverse the mineral oil. The event signature is the same as for previous $\nu$ and  $\nubar$ NCE cross section analyses performed by MB~\cite{AguilarArevalo:2010cx,Aguilar-Arevalo:2013nkf}. 

\noindent \textbf{Analysis ---}
A DM-candidate event sample was selected from the off-target data with selection criteria (``cuts'') following the previous MB $\nubar$-NCE analysis~\cite{Aguilar-Arevalo:2013nkf} and a reconstructed nucleon kinetic energy of $35<T_p<600$~MeV.  This procedure requires exactly one time-cluster of hits coincident with the beam and with a time and spatial distribution consistent with a single nucleon and no pions.  This selection, along with the requirement of no activity in the veto, minimizes contamination from beam-unrelated (cosmic) backgrounds and non-NCE beam-related backgrounds.  Neutrino-induced NCE events are an irreducible background to this analysis; they must be estimated and subtracted. To better constrain the off-target neutrino flux and, therefore, the neutrino-induced backgrounds, the set of off-target CCQE events mentioned above was selected following cuts developed for our $\nu$-mode CCQE cross section measurement~\cite{AguilarArevalo:2010zc}. 

Because the NCE and CCQE cross sections are not known \textit{a priori} independently of MB data, two other large samples ($\approx $100k events), with the same NCE and CCQE cuts as for the beam-off-target set, were extracted from previously-collected $\nu$-mode data.  The DM-candidate sample contains any $\chi N$ scattering events while the three other ``constraint'' samples serve to constrain the event rate for an improved estimate of beam-related backgrounds. It should be noted that the events passing the selection cuts are not purely NCE and CCQE at the vertex level but are more acurately labeled ``NC0$\pi$''and ``CC0$\pi$'' because of processes like pion-production combined with pion absorption in the nucleus or scattering via multinucleon processes~\cite{Katori:2016yel}. 
 
A detector simulation, developed and tuned for previous MB analyses, but with the new off-target neutrino flux, was used to predict the event rates for these neutrino-induced processes including those involving pion absorption.  The simulation predicts that the NC0$\pi$ (CC0$\pi$) samples consist of  77\% (84\%) true NCE (CCQE) events but the analysis does not depend strongly on those values since the constraint samples determine the effective cross sections. The simulation is also used to determine the DM event efficiency and related errors including correlations~\cite{AguilarArevalo:2010zc,AguilarArevalo:2010cx}.
 
\begin{figure}
\includegraphics[width=0.99\columnwidth]{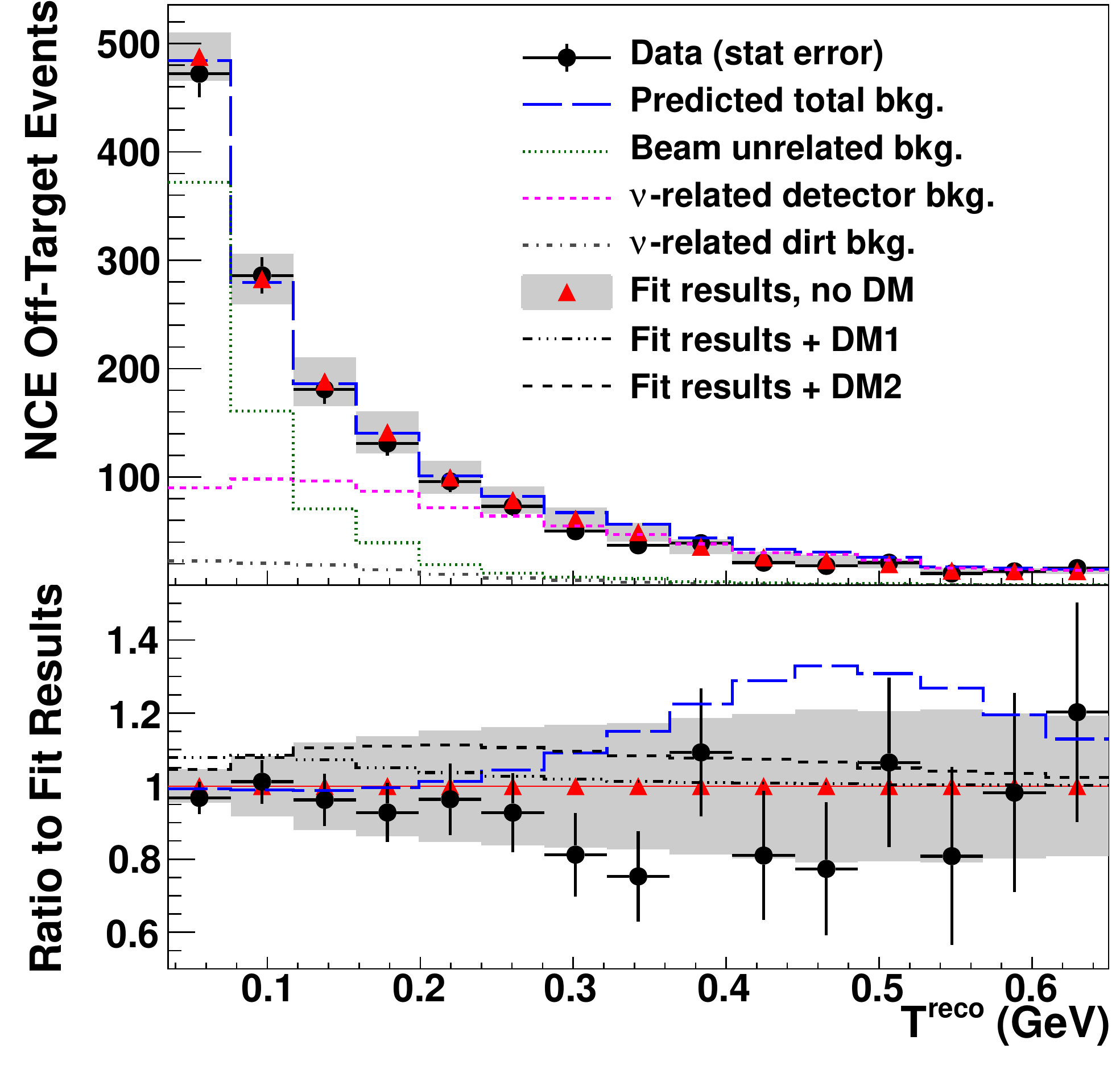}
\caption{\label{fig:NCETN} Reconstructed nucleon kinetic energy distribution for DM candidate events with the experimental data are shown as circles with statistical error bars. The predicted backgrounds are shown as lines and the results from a background-only fit to the combined data set are shown as triangles with error boxes.   The bottom plot shows the data and unconstrained background-only prediction together with example DM signals as a ratio to the background-only fit.  The example signals are the 90\% confidence-limit solutions at the best-fit point (DM1, $m_V = 10~\mathrm{MeV}, m_\chi = 1~\mathrm{MeV}, \epalp=8.1\times 10^{-14}$) and the most-sensitive point (DM2, $m_V = 769~\mathrm{MeV}, m_\chi = 381~\mathrm{MeV}, \epalp=1.3\times 10^{-14}$). }
\end{figure}

The nucleon reconstructed kinetic energy distribution for DM candidate events is shown in Fig.~\ref{fig:NCETN} and the integrated event totals are summarized in Tab.~\ref{tab:NCEoffEvents}.  The background predictions are determined through both measurement and simulations. The beam-unrelated background is measured in out-of-beam, 19.2~$\mu$s-duration windows taken at 10-15~Hz interspersed with the beam-on data-collection windows.   The same cuts are then applied to this sample for an estimate of the number of beam-unrelated events passing cuts in the beam-on sample.   

The beam-related detector background is dominated by NCE events originating within the detector volume and are estimated using the experimental simulation.  Beam-related ``dirt'' backgrounds arise mainly from neutrino-induced neutrons created outside the detector, passing into the main detector volume, and satisfying the event selection.   All of these beam-related background processes have been measured in various MB data sets and then used as input to the simulations.
 
\begin{table}
  \caption{\label{tab:NCEoffEvents} Number of selected data events with predicted backgrounds.}
  \begin{ruledtabular}
    \begin{tabular}{p{2.7in}r@{ $\pm$}r}
      background source & \multicolumn{2}{c}{events} \\
      \hline
	  beam-unrelated \hfill          & 697 & 11 \\
	  beam-related, detector        & 775 &  454  \\
	  beam-related, dirt                & 107 & 81 \\
	  \hline\hline
	  total estimated background  & 1579   &  529 \\ 
      constrained-fit background     & 1548  & 198 \\ 
	  \hline
	  data events                         & 1465 & 38  \\
    \end{tabular}
  \end{ruledtabular}
\end{table}

As seen in Table~\ref{tab:NCEoffEvents}, the error on the beam-unrelated background is small and due to statistical error in the large beam-off sample; the systematic error is negligible.   The largest errors are those on the beam-related background estimates which originate from uncertainties on the neutrino flux, NCE cross section model, and detector response.  Correlated errors between different energy bins and event samples are also calculated.   The resulting error using this procedure is $34\%$ of the estimated background while the statistical error on the data is $3\%$. This measurement is systematic-error limited.  
  
However, this systematic error was reduced substantially via a combined fit of the DM-candidate sample together with the three constraint samples described above.  Effectively, the off-target CCQE sample determines the off-target flux with errors smaller than those resulting from the simulation procedure.  Similarly,  the NCE sample from $\nu$-mode determines the event rate for neutrino background processes with reduced errors.  As shown in Table~\ref{tab:NCEoffEvents}, the error on the background is reduced from $34\%$ to $13\%$ with this ``constrained-fit'' procedure.  The energy distribution of predicted background events resulting from this fit is shown in Fig.~\ref{fig:NCETN} with the reduced errors. 

A signal for DM would appear as an excess of events above background such as that shown for two example DM parameter sets in Fig.~\ref{fig:NCETN}.  The data show no significant excess of events over the background prediction and may be used to set limits on the vector portal DM model parameters. 

 A background-only fit on the full data set, consisting of DM candidate events and constraint samples, was the first step in the procedure.  In order to allow some adjustment of the underlying background distributions within errors, six ``nuisance'' parameters were introduced: one scale factor each for the $\nu$-mode and off-target neutrino fluxes, and four parameters to adjust the NCE cross section.   As can be seen in~\cite{AguilarArevalo:2010cx, Aguilar-Arevalo:2013nkf} the simulation overpredicts the NCE data at higher nucleon energy and may be due to an overestimate of pion background channels.  These nuisance parameters, consisting of an overall normalization factor together with a subtracted Gaussian, correct this.  The predicted backgrounds, adjusted by the nuisance parameters, were then fit to the four data samples in a total of 80 bins of calculated 4-momentum transfer using a log-likelihood function constructed with the complete and correlated ($80\times80$) error matrix.  The resulting $\chi^2$ was 48.1/74 giving an upper tail probability of 97\%, reflecting fairly conservative errors, which is not surprising as the simulations have been pre-tuned somewhat on existing data samples.  

 The next step was to use a fixed-target DM simulation~\cite{deNiverville:2016rqh} to generate predicted energy and position distributions of expected $\chi N$ scattering events in the MB detector for a particular set of DM parameters.  The simulation, based on the model described in detail in Ref.~\cite{deNiverville:2016rqh}, calculates rates for DM production and interactions in the detector as described in the introduction above.  The attenuation of the $\chi$ flux in the beam dump and earth shield was calculated and is negligible for the model parameters considered here.  The kinematic distributions of the particles involved for these mechanisms were obtained from the beam simulations.  The energy distribution of the DM scattered nucleons from the DM simulation was used as input to the MB detector simulation which then could be used to calculate event efficiencies and generate a predicted nucleon energy distribution.   In practice, since $\chi N$ events have the same final-state signature as the NCE sample, existing simulation samples were used for a $\chi N$ sample with an event-weight scaling based on the scattered nucleon energy.   Only true NCE events were used for the DM signal.  This is equivalent to assuming no DM interactions via resonant events and will result in a more conservative limit.  The efficiency for a DM scattering event to be detected in this analysis is $\approx$ 35\% for nucleon kinetic energy above $\approx 150$~MeV but falls rapidly to $< 1\%$ at 50~MeV.  In addition, the nucleons in carbon are subject to binding energy and final-state interactions further reducing the efficiency.  The DM simulation of \cite{deNiverville:2016rqh} does not include corrections for bound nucleons so they were applied using an effective efficiency calculated from the MB simulation which does account for those effects~\cite{AguilarArevalo:2010zc}. 

\begin{figure}
\includegraphics[width=0.99\columnwidth]{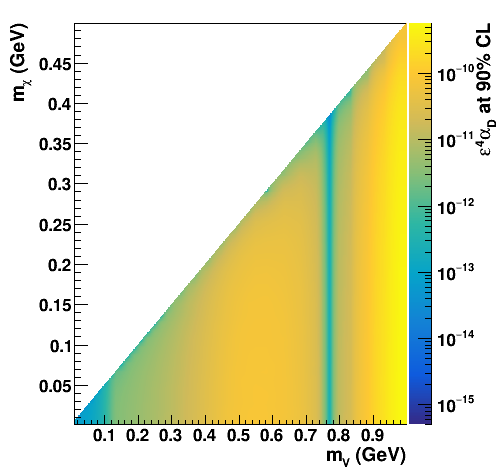}
\caption{\label{fig:cl} The $\epsilon^4\alpha_D$ 90\% confidence limits for $0.01<m_V<1$~GeV and $m_V>2m_\chi$ using the vector portal DM model.}
\end{figure}

 The procedure results in a set of predicted $\chi N$ signal events for each set of $\epsilon^4\alpha_D$, $m_V$, and $m_\chi$.  The number of predicted events simply scales with the $\epsilon^4\alpha_D$ parameter, while the nucleon energy distribution changes shape with each $m_V$ and $m_\chi$.  These DM simulation results were then combined with the components described in the background-only fit above and subjected to a frequentist confidence limit (CL) method developed previously for the MB $\nu$ and $\nubar$ oscillations analyses~\cite{AlexisThesis,Aguilar-Arevalo:2013pmq}.  The procedure determines the 90\% CL $\epalp$  value within this vector portal DM model and allowed by this experimental data set for a given $m_V,m_\chi$ pair with $0.01<m_\chi<0.5$~GeV, $m_V>2m_\chi$.  These results (Fig.~\ref{fig:cl}) provide the best sensitivity of $\epalp<1.2\times10^{-14}$ at $m_V\approx775$~MeV, near the $\rho$ and $\omega$ masses.
 
\begin{figure}
\includegraphics[width=0.99\columnwidth]{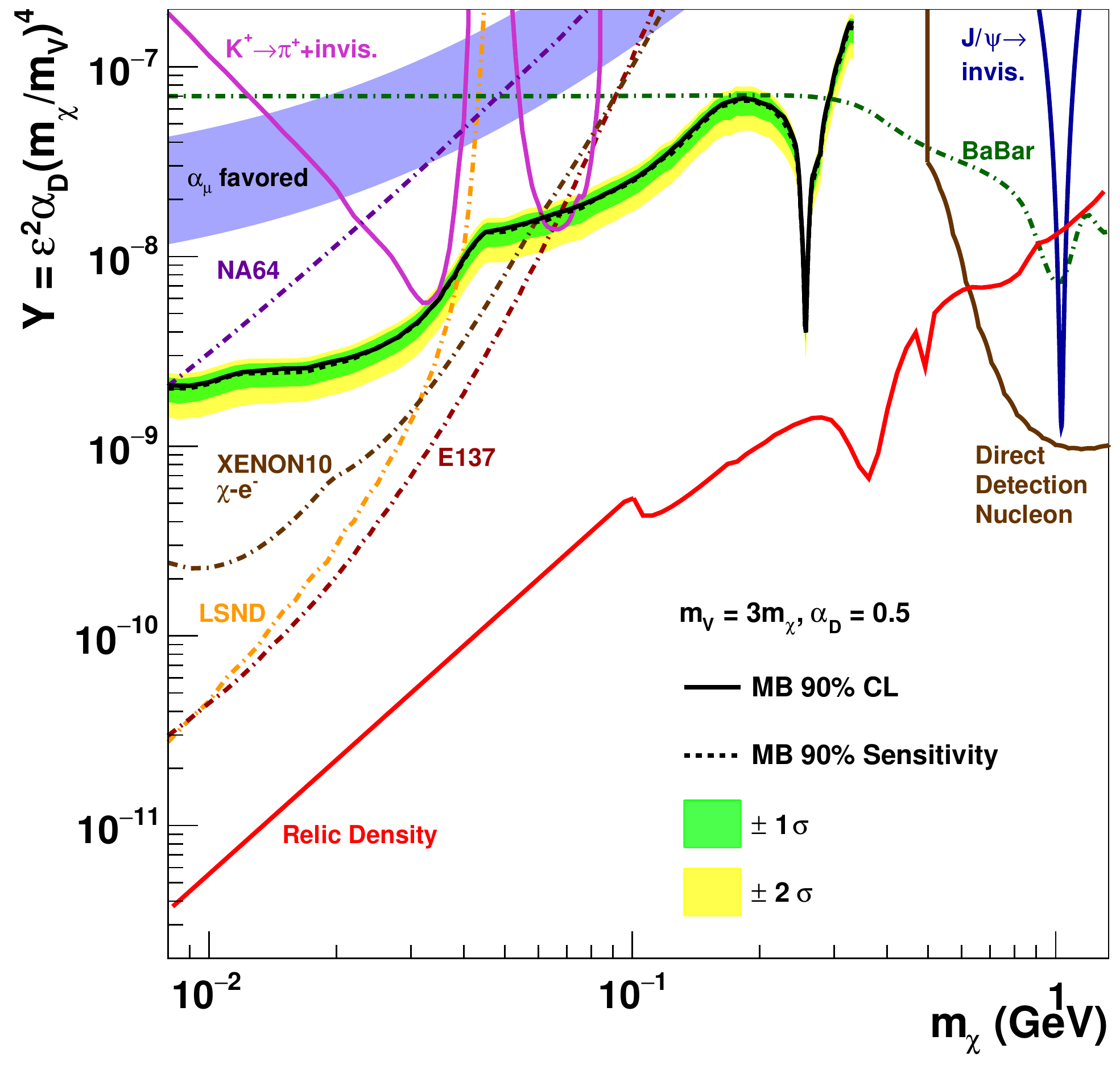}
\caption{\label{fig:clKappa} Confidence limits and sensitivities with $1,2\sigma$ errors resulting from this analysis compared to other experimental results
~\cite{Batell:2014mga,deNiverville:2011it,Pospelov:2008zw,Fayet:2007ua,Essig:2013vha,Dobrescu:2014ita,Agnese:2015nto,Angloher:2015ewa,Banerjee:2016tad,Essig:2012yx}.
Limits from experiments that assume DM coupling to quarks/nucleons, including this result, are shown as solid lines while those that require DM coupling to electrons are shown as dot-dashed lines. 
The favored parameters for this model to account for the observed relic DM density~\cite{deNiverville:2011it} are shown as the lowest solid line.}
\end{figure}

\noindent \textbf{Conclusions ---}
This analysis determines the 90\% CL value for the combination $\epalp$.  Using conventional choices for the other DM parameters allows comparisons of experiments employing different methods in a shared parameter space.  In Fig.~\ref{fig:clKappa}, with $m_V=3m_\chi$ and $\alpha_D=0.5$, the 90\% CL values for the dimensionless DM annihilation cross section parameter $Y=\epsilon^2\alpha_D(m_\chi/m_V)^4$  may be plotted for this result and compared to different experimental exclusion regions.  The choice of $\alpha_D = 0.5$ is compatible with the bounds derived in Ref.~\cite{Davoudiasl:2015hxa} based on the running of the dark gauge coupling. However, it is important to note that the $\chi$ yield scales as $\epsilon^4 \alpha_D$. Thus for sufficiently small values of  $\alpha_D$ the limits from other probes such as BaBar\cite{Essig:2013vha} will be stronger.
With these DM parameter combinations, this result has expanded the search for DM to $m_\chi$ values 2 orders of magnitude smaller than nucleon direct detection DM experiments and has excluded a vector mediator particle solution to the $g-2$ anomaly~\cite{Pospelov:2008zw,Fayet:2007ua}.   Within the context of the vector portal DM model and the chosen parameter constraints, this result sets the most stringent limits on DM in the range $0.08<m_\chi<0.3$~GeV and, in a model where the DM does not couple to electrons~\cite{Batell:2014yra}, this limit is extended down to $m_\chi \approx 0.01$~GeV.

\noindent \textit{Acknowlegments}
This work was supported by the U.S. Department ofEnergy; the U.S. National Science Foundation; Los Alamos National Laboratory, U.S., Science and Technology Facilities Council, UK; Consejo Nacional de Ciencia y Tecnologa, Mexico. We thank Fermilab Accelerator Divison for the work to recongure, operate, and understand the off-target beam. Fermilab is operated by Fermi Research Alliance, LLC under Contract No. De-AC02-07CH11359 with the US DOE.

\input{MBndm.bbl}

\end{document}

%% file: MBndmAuthors.tex
\newcommand{\alabama}{University of Alabama, Tuscaloosa, AL 35487}
\newcommand{\anl}{Argonne National Laboratory, Argonne, IL 60439}
\newcommand{\chicago}{University of Chicago, Chicago, IL, 60637}
\newcommand{\cincinnati}{University of Cincinnati, Cincinnati, OH 45221}
\newcommand{\columbia}{Columbia University, New York, NY 10027}
\newcommand{\ctpu}{Center for Theoretical Physics of the Universe, Institute for Basic Science (IBS), Daejeon, 34051, Korea}
\newcommand{\fermi}{Fermi National Accelerator Laboratory, Batavia, IL 60510}
\newcommand{\florida}{University of Florida, Gainesville, FL 32611}
\newcommand{\indiana}{Indiana University, Bloomington, IN 47405}
\newcommand{\lanl}{Los Alamos National Laboratory, Los Alamos, NM 87545}
\newcommand{\michigan}{University of Michigan, Ann Arbor, MI 48111}
\newcommand{\mexico}{Instituto de Ciencias Nucleares, Universidad Nacional Aut\'onoma de M\'exico, Mexico City 04510, Mexico}
\newcommand{\nmsu}{New Mexico State University, Las Cruces, NM 88003}
\newcommand{\pittsburgh}{University of Pittsburgh, Pittsburgh, PA 15260, USA}
\newcommand{\qmery}{Queen Mary University of London, London, E1 4NS, UK}
\newcommand{\smary}{Saint Mary's University of Minnesota, Winona, MN 55987}
\newcommand{\uta}{University of Texas (Arlington), Arlington, TX 76019}


\author{A.A.~Aguilar-Arevalo}\affiliation{\mexico}
\author{M.~Backfish}\affiliation{\fermi}
\author{A.~Bashyal}\affiliation{\uta}
\author{B.~Batell}\affiliation{\pittsburgh}
\author{B.C.~Brown}\affiliation{\fermi}
\author{R.~Carr}\affiliation{\columbia}
\author{A.~Chatterjee}\affiliation{\uta}
\author{R.L.~Cooper}\affiliation{\indiana}\affiliation{\nmsu}
\author{P.~deNiverville}\affiliation{\ctpu}
\author{R.~Dharmapalan}\affiliation{\anl}
\author{Z.~Djurcic}\affiliation{\anl}
\author{R.~Ford}\affiliation{\fermi}
\author{F.G.~Garcia}\affiliation{\fermi}
\author{G.T.~Garvey}\affiliation{\lanl}
\author{J.~Grange}\affiliation{\anl}\affiliation{\florida}
\author{J.A.~Green}\affiliation{\lanl}
\author{W.~Huelsnitz}\affiliation{\lanl}
\author{I.L.~de Icaza Astiz}\affiliation{\mexico}
\author{G.~Karagiorgi}\affiliation{\columbia}
\author{T.~Katori}\affiliation{\qmery}
\author{W.~Ketchum}\affiliation{\lanl}
\author{T.~Kobilarcik}\affiliation{\fermi}
\author{Q.~Liu}\affiliation{\lanl}
\author{W.C.~Louis}\affiliation{\lanl}
\author{W.~Marsh}\affiliation{\fermi}
\author{C.D.~Moore}\affiliation{\fermi}
\author{G.B.~Mills}\affiliation{\lanl}
\author{J.~Mirabal}\affiliation{\lanl}
\author{P.~Nienaber}\affiliation{\smary}
\author{Z.~Pavlovic}\affiliation{\lanl}
\author{D.~Perevalov}\affiliation{\fermi}
\author{H.~Ray}\affiliation{\florida}
\author{B.P.~Roe}\affiliation{\michigan}
\author{M.H.~Shaevitz}\affiliation{\columbia}
\author{S.~Shahsavarani}\affiliation{\uta}
\author{I.~Stancu}\affiliation{\alabama}
\author{R.~Tayloe}\affiliation{\indiana}
\author{C.~Taylor}\affiliation{\lanl}
\author{R.T.~Thornton}\affiliation{\indiana}
\author{R.~Van de Water}\affiliation{\lanl}
\author{W. Wester}\affiliation{\fermi}
\author{D.H.~White}\affiliation{\lanl}
\author{J. Yu}\affiliation{\uta}

%% file: MBndm.bbl
%

%% file: MBndm.bbl
\begin{thebibliography}{37}%
\makeatletter
\providecommand \@ifxundefined [1]{%
 \@ifx{#1\undefined}
}%
\providecommand \@ifnum [1]{%
 \ifnum #1\expandafter \@firstoftwo
 \else \expandafter \@secondoftwo
 \fi
}%
\providecommand \@ifx [1]{%
 \ifx #1\expandafter \@firstoftwo
 \else \expandafter \@secondoftwo
 \fi
}%
\providecommand \natexlab [1]{#1}%
\providecommand \enquote  [1]{``#1''}%
\providecommand \bibnamefont  [1]{#1}%
\providecommand \bibfnamefont [1]{#1}%
\providecommand \citenamefont [1]{#1}%
\providecommand \href@noop [0]{\@secondoftwo}%
\providecommand \href [0]{\begingroup \@sanitize@url \@href}%
\providecommand \@href[1]{\@@startlink{#1}\@@href}%
\providecommand \@@href[1]{\endgroup#1\@@endlink}%
\providecommand \@sanitize@url [0]{\catcode `\\12\catcode `\$12\catcode
  `\&12\catcode `\#12\catcode `\^12\catcode `\_12\catcode `\%12\relax}%
\providecommand \@@startlink[1]{}%
\providecommand \@@endlink[0]{}%
\providecommand \url  [0]{\begingroup\@sanitize@url \@url }%
\providecommand \@url [1]{\endgroup\@href {#1}{\urlprefix }}%
\providecommand \urlprefix  [0]{URL }%
\providecommand \Eprint [0]{\href }%
\providecommand \doibase [0]{http://dx.doi.org/}%
\providecommand \selectlanguage [0]{\@gobble}%
\providecommand \bibinfo  [0]{\@secondoftwo}%
\providecommand \bibfield  [0]{\@secondoftwo}%
\providecommand \translation [1]{[#1]}%
\providecommand \BibitemOpen [0]{}%
\providecommand \bibitemStop [0]{}%
\providecommand \bibitemNoStop [0]{.\EOS\space}%
\providecommand \EOS [0]{\spacefactor3000\relax}%
\providecommand \BibitemShut  [1]{\csname bibitem#1\endcsname}%
\let\auto@bib@innerbib\@empty
\bibitem [{\citenamefont {Patrignani}\ \emph {et~al.}(2016)\citenamefont
  {Patrignani} \emph {et~al.}}]{Olive:2016xmw}%
  \BibitemOpen
  \bibfield  {author} {\bibinfo {author} {\bibfnamefont {C.}~\bibnamefont
  {Patrignani}} \emph {et~al.} (\bibinfo {collaboration} {Particle Data
  Group}),\ }\href {\doibase 10.1088/1674-1137/40/10/100001} {\bibfield
  {journal} {\bibinfo  {journal} {Chin. Phys.}\ }\textbf {\bibinfo {volume}
  {C40}},\ \bibinfo {pages} {100001} (\bibinfo {year} {2016})}\BibitemShut
  {NoStop}%
\bibitem [{\citenamefont {Alexander}\ \emph {et~al.}(2016)\citenamefont
  {Alexander} \emph {et~al.}}]{Alexander:2016aln}%
  \BibitemOpen
  \bibfield  {author} {\bibinfo {author} {\bibfnamefont {J.}~\bibnamefont
  {Alexander}} \emph {et~al.}\ }(\bibinfo {year} {2016})\ \Eprint
  {http://arxiv.org/abs/1608.08632} {arXiv:1608.08632 [hep-ph]} \BibitemShut
  {NoStop}%
\bibitem [{\citenamefont {Batell}\ \emph {et~al.}(2009)\citenamefont {Batell},
  \citenamefont {Pospelov},\ and\ \citenamefont {Ritz}}]{Batell:2009di}%
  \BibitemOpen
  \bibfield  {author} {\bibinfo {author} {\bibfnamefont {B.}~\bibnamefont
  {Batell}}, \bibinfo {author} {\bibfnamefont {M.}~\bibnamefont {Pospelov}}, \
  and\ \bibinfo {author} {\bibfnamefont {A.}~\bibnamefont {Ritz}},\ }\href
  {\doibase 10.1103/PhysRevD.80.095024} {\bibfield  {journal} {\bibinfo
  {journal} {Phys.Rev.}\ }\textbf {\bibinfo {volume} {D80}},\ \bibinfo {pages}
  {095024} (\bibinfo {year} {2009})},\ \Eprint {http://arxiv.org/abs/0906.5614}
  {arXiv:0906.5614 [hep-ph]} \BibitemShut {NoStop}%
\bibitem [{\citenamefont {deNiverville}\ \emph {et~al.}(2011)\citenamefont
  {deNiverville}, \citenamefont {Pospelov},\ and\ \citenamefont
  {Ritz}}]{deNiverville:2011it}%
  \BibitemOpen
  \bibfield  {author} {\bibinfo {author} {\bibfnamefont {P.}~\bibnamefont
  {deNiverville}}, \bibinfo {author} {\bibfnamefont {M.}~\bibnamefont
  {Pospelov}}, \ and\ \bibinfo {author} {\bibfnamefont {A.}~\bibnamefont
  {Ritz}},\ }\href {\doibase 10.1103/PhysRevD.84.075020} {\bibfield  {journal}
  {\bibinfo  {journal} {Phys.Rev.}\ }\textbf {\bibinfo {volume} {D84}},\
  \bibinfo {pages} {075020} (\bibinfo {year} {2011})},\ \Eprint
  {http://arxiv.org/abs/1107.4580} {arXiv:1107.4580 [hep-ph]} \BibitemShut
  {NoStop}%
\bibitem [{\citenamefont {deNiverville}\ \emph {et~al.}(2012)\citenamefont
  {deNiverville}, \citenamefont {McKeen},\ and\ \citenamefont
  {Ritz}}]{deNiverville:2012ij}%
  \BibitemOpen
  \bibfield  {author} {\bibinfo {author} {\bibfnamefont {P.}~\bibnamefont
  {deNiverville}}, \bibinfo {author} {\bibfnamefont {D.}~\bibnamefont
  {McKeen}}, \ and\ \bibinfo {author} {\bibfnamefont {A.}~\bibnamefont
  {Ritz}},\ }\href {\doibase 10.1103/PhysRevD.86.035022} {\bibfield  {journal}
  {\bibinfo  {journal} {Phys.Rev.}\ }\textbf {\bibinfo {volume} {D86}},\
  \bibinfo {pages} {035022} (\bibinfo {year} {2012})},\ \Eprint
  {http://arxiv.org/abs/1205.3499} {arXiv:1205.3499 [hep-ph]} \BibitemShut
  {NoStop}%
\bibitem [{\citenamefont {Dharmapalan}\ \emph {et~al.}(2012)\citenamefont
  {Dharmapalan} \emph {et~al.}}]{Dharmapalan:2012xp}%
  \BibitemOpen
  \bibfield  {author} {\bibinfo {author} {\bibfnamefont {R.}~\bibnamefont
  {Dharmapalan}} \emph {et~al.} (\bibinfo {collaboration} {MiniBooNE}),\
  }\href@noop {} {\  (\bibinfo {year} {2012})},\ \Eprint
  {http://arxiv.org/abs/1211.2258} {arXiv:1211.2258 [hep-ex]} \BibitemShut
  {NoStop}%
\bibitem [{\citenamefont {Izaguirre}\ \emph {et~al.}(2013)\citenamefont
  {Izaguirre}, \citenamefont {Krnjaic}, \citenamefont {Schuster},\ and\
  \citenamefont {Toro}}]{Izaguirre:2013uxa}%
  \BibitemOpen
  \bibfield  {author} {\bibinfo {author} {\bibfnamefont {E.}~\bibnamefont
  {Izaguirre}}, \bibinfo {author} {\bibfnamefont {G.}~\bibnamefont {Krnjaic}},
  \bibinfo {author} {\bibfnamefont {P.}~\bibnamefont {Schuster}}, \ and\
  \bibinfo {author} {\bibfnamefont {N.}~\bibnamefont {Toro}},\ }\href {\doibase
  10.1103/PhysRevD.88.114015} {\bibfield  {journal} {\bibinfo  {journal}
  {Phys.Rev.}\ }\textbf {\bibinfo {volume} {D88}},\ \bibinfo {pages} {114015}
  (\bibinfo {year} {2013})},\ \Eprint {http://arxiv.org/abs/1307.6554}
  {arXiv:1307.6554 [hep-ph]} \BibitemShut {NoStop}%
\bibitem [{\citenamefont {Izaguirre}\ \emph
  {et~al.}(2015{\natexlab{a}})\citenamefont {Izaguirre}, \citenamefont
  {Krnjaic}, \citenamefont {Schuster},\ and\ \citenamefont
  {Toro}}]{Izaguirre:2014bca}%
  \BibitemOpen
  \bibfield  {author} {\bibinfo {author} {\bibfnamefont {E.}~\bibnamefont
  {Izaguirre}}, \bibinfo {author} {\bibfnamefont {G.}~\bibnamefont {Krnjaic}},
  \bibinfo {author} {\bibfnamefont {P.}~\bibnamefont {Schuster}}, \ and\
  \bibinfo {author} {\bibfnamefont {N.}~\bibnamefont {Toro}},\ }\href {\doibase
  10.1103/PhysRevD.91.094026} {\bibfield  {journal} {\bibinfo  {journal} {Phys.
  Rev.}\ }\textbf {\bibinfo {volume} {D91}},\ \bibinfo {pages} {094026}
  (\bibinfo {year} {2015}{\natexlab{a}})},\ \Eprint
  {http://arxiv.org/abs/1411.1404} {arXiv:1411.1404 [hep-ph]} \BibitemShut
  {NoStop}%
\bibitem [{\citenamefont {Izaguirre}\ \emph {et~al.}(2014)\citenamefont
  {Izaguirre}, \citenamefont {Krnjaic}, \citenamefont {Schuster},\ and\
  \citenamefont {Toro}}]{Izaguirre:2014dua}%
  \BibitemOpen
  \bibfield  {author} {\bibinfo {author} {\bibfnamefont {E.}~\bibnamefont
  {Izaguirre}}, \bibinfo {author} {\bibfnamefont {G.}~\bibnamefont {Krnjaic}},
  \bibinfo {author} {\bibfnamefont {P.}~\bibnamefont {Schuster}}, \ and\
  \bibinfo {author} {\bibfnamefont {N.}~\bibnamefont {Toro}},\ }\href {\doibase
  10.1103/PhysRevD.90.014052} {\bibfield  {journal} {\bibinfo  {journal} {Phys.
  Rev.}\ }\textbf {\bibinfo {volume} {D90}},\ \bibinfo {pages} {014052}
  (\bibinfo {year} {2014})},\ \Eprint {http://arxiv.org/abs/1403.6826}
  {arXiv:1403.6826 [hep-ph]} \BibitemShut {NoStop}%
\bibitem [{\citenamefont {Batell}\ \emph
  {et~al.}(2014{\natexlab{a}})\citenamefont {Batell}, \citenamefont
  {deNiverville}, \citenamefont {McKeen}, \citenamefont {Pospelov},\ and\
  \citenamefont {Ritz}}]{Batell:2014yra}%
  \BibitemOpen
  \bibfield  {author} {\bibinfo {author} {\bibfnamefont {B.}~\bibnamefont
  {Batell}}, \bibinfo {author} {\bibfnamefont {P.}~\bibnamefont
  {deNiverville}}, \bibinfo {author} {\bibfnamefont {D.}~\bibnamefont
  {McKeen}}, \bibinfo {author} {\bibfnamefont {M.}~\bibnamefont {Pospelov}}, \
  and\ \bibinfo {author} {\bibfnamefont {A.}~\bibnamefont {Ritz}},\ }\href
  {\doibase 10.1103/PhysRevD.90.115014} {\bibfield  {journal} {\bibinfo
  {journal} {Phys. Rev.}\ }\textbf {\bibinfo {volume} {D90}},\ \bibinfo {pages}
  {115014} (\bibinfo {year} {2014}{\natexlab{a}})},\ \Eprint
  {http://arxiv.org/abs/1405.7049} {arXiv:1405.7049 [hep-ph]} \BibitemShut
  {NoStop}%
\bibitem [{\citenamefont {Batell}\ \emph
  {et~al.}(2014{\natexlab{b}})\citenamefont {Batell}, \citenamefont {Essig},\
  and\ \citenamefont {Surujon}}]{Batell:2014mga}%
  \BibitemOpen
  \bibfield  {author} {\bibinfo {author} {\bibfnamefont {B.}~\bibnamefont
  {Batell}}, \bibinfo {author} {\bibfnamefont {R.}~\bibnamefont {Essig}}, \
  and\ \bibinfo {author} {\bibfnamefont {Z.}~\bibnamefont {Surujon}},\ }\href
  {\doibase 10.1103/PhysRevLett.113.171802} {\bibfield  {journal} {\bibinfo
  {journal} {Phys.Rev.Lett.}\ }\textbf {\bibinfo {volume} {113}},\ \bibinfo
  {pages} {171802} (\bibinfo {year} {2014}{\natexlab{b}})},\ \Eprint
  {http://arxiv.org/abs/1406.2698} {arXiv:1406.2698 [hep-ph]} \BibitemShut
  {NoStop}%
\bibitem [{\citenamefont {Dobrescu}\ and\ \citenamefont
  {Frugiuele}(2015)}]{Dobrescu:2014ita}%
  \BibitemOpen
  \bibfield  {author} {\bibinfo {author} {\bibfnamefont {B.~A.}\ \bibnamefont
  {Dobrescu}}\ and\ \bibinfo {author} {\bibfnamefont {C.}~\bibnamefont
  {Frugiuele}},\ }\href {\doibase 10.1007/JHEP02(2015)019} {\bibfield
  {journal} {\bibinfo  {journal} {JHEP}\ }\textbf {\bibinfo {volume} {02}},\
  \bibinfo {pages} {019} (\bibinfo {year} {2015})},\ \Eprint
  {http://arxiv.org/abs/1410.1566} {arXiv:1410.1566 [hep-ph]} \BibitemShut
  {NoStop}%
\bibitem [{\citenamefont {Soper}\ \emph {et~al.}(2014)\citenamefont {Soper},
  \citenamefont {Spannowsky}, \citenamefont {Wallace},\ and\ \citenamefont
  {Tait}}]{Soper:2014ska}%
  \BibitemOpen
  \bibfield  {author} {\bibinfo {author} {\bibfnamefont {D.~E.}\ \bibnamefont
  {Soper}}, \bibinfo {author} {\bibfnamefont {M.}~\bibnamefont {Spannowsky}},
  \bibinfo {author} {\bibfnamefont {C.~J.}\ \bibnamefont {Wallace}}, \ and\
  \bibinfo {author} {\bibfnamefont {T.~M.~P.}\ \bibnamefont {Tait}},\ }\href
  {\doibase 10.1103/PhysRevD.90.115005} {\bibfield  {journal} {\bibinfo
  {journal} {Phys. Rev.}\ }\textbf {\bibinfo {volume} {D90}},\ \bibinfo {pages}
  {115005} (\bibinfo {year} {2014})},\ \Eprint {http://arxiv.org/abs/1407.2623}
  {arXiv:1407.2623 [hep-ph]} \BibitemShut {NoStop}%
\bibitem [{\citenamefont {Kahn}\ \emph {et~al.}(2015)\citenamefont {Kahn},
  \citenamefont {Krnjaic}, \citenamefont {Thaler},\ and\ \citenamefont
  {Toups}}]{Kahn:2014sra}%
  \BibitemOpen
  \bibfield  {author} {\bibinfo {author} {\bibfnamefont {Y.}~\bibnamefont
  {Kahn}}, \bibinfo {author} {\bibfnamefont {G.}~\bibnamefont {Krnjaic}},
  \bibinfo {author} {\bibfnamefont {J.}~\bibnamefont {Thaler}}, \ and\ \bibinfo
  {author} {\bibfnamefont {M.}~\bibnamefont {Toups}},\ }\href {\doibase
  10.1103/PhysRevD.91.055006} {\bibfield  {journal} {\bibinfo  {journal} {Phys.
  Rev.}\ }\textbf {\bibinfo {volume} {D91}},\ \bibinfo {pages} {055006}
  (\bibinfo {year} {2015})},\ \Eprint {http://arxiv.org/abs/1411.1055}
  {arXiv:1411.1055 [hep-ph]} \BibitemShut {NoStop}%
\bibitem [{\citenamefont {Izaguirre}\ \emph
  {et~al.}(2015{\natexlab{b}})\citenamefont {Izaguirre}, \citenamefont
  {Krnjaic}, \citenamefont {Schuster},\ and\ \citenamefont
  {Toro}}]{Izaguirre:2015yja}%
  \BibitemOpen
  \bibfield  {author} {\bibinfo {author} {\bibfnamefont {E.}~\bibnamefont
  {Izaguirre}}, \bibinfo {author} {\bibfnamefont {G.}~\bibnamefont {Krnjaic}},
  \bibinfo {author} {\bibfnamefont {P.}~\bibnamefont {Schuster}}, \ and\
  \bibinfo {author} {\bibfnamefont {N.}~\bibnamefont {Toro}},\ }\href {\doibase
  10.1103/PhysRevLett.115.251301} {\bibfield  {journal} {\bibinfo  {journal}
  {Phys. Rev. Lett.}\ }\textbf {\bibinfo {volume} {115}},\ \bibinfo {pages}
  {251301} (\bibinfo {year} {2015}{\natexlab{b}})},\ \Eprint
  {http://arxiv.org/abs/1505.00011} {arXiv:1505.00011 [hep-ph]} \BibitemShut
  {NoStop}%
\bibitem [{\citenamefont {deNiverville}\ \emph {et~al.}(2015)\citenamefont
  {deNiverville}, \citenamefont {Pospelov},\ and\ \citenamefont
  {Ritz}}]{deNiverville:2015mwa}%
  \BibitemOpen
  \bibfield  {author} {\bibinfo {author} {\bibfnamefont {P.}~\bibnamefont
  {deNiverville}}, \bibinfo {author} {\bibfnamefont {M.}~\bibnamefont
  {Pospelov}}, \ and\ \bibinfo {author} {\bibfnamefont {A.}~\bibnamefont
  {Ritz}},\ }\href {\doibase 10.1103/PhysRevD.92.095005} {\bibfield  {journal}
  {\bibinfo  {journal} {Phys. Rev.}\ }\textbf {\bibinfo {volume} {D92}},\
  \bibinfo {pages} {095005} (\bibinfo {year} {2015})},\ \Eprint
  {http://arxiv.org/abs/1505.07805} {arXiv:1505.07805 [hep-ph]} \BibitemShut
  {NoStop}%
\bibitem [{\citenamefont {Izaguirre}\ \emph
  {et~al.}(2015{\natexlab{c}})\citenamefont {Izaguirre}, \citenamefont
  {Krnjaic},\ and\ \citenamefont {Pospelov}}]{Izaguirre:2015pva}%
  \BibitemOpen
  \bibfield  {author} {\bibinfo {author} {\bibfnamefont {E.}~\bibnamefont
  {Izaguirre}}, \bibinfo {author} {\bibfnamefont {G.}~\bibnamefont {Krnjaic}},
  \ and\ \bibinfo {author} {\bibfnamefont {M.}~\bibnamefont {Pospelov}},\
  }\href {\doibase 10.1103/PhysRevD.92.095014} {\bibfield  {journal} {\bibinfo
  {journal} {Phys. Rev.}\ }\textbf {\bibinfo {volume} {D92}},\ \bibinfo {pages}
  {095014} (\bibinfo {year} {2015}{\natexlab{c}})},\ \Eprint
  {http://arxiv.org/abs/1507.02681} {arXiv:1507.02681 [hep-ph]} \BibitemShut
  {NoStop}%
\bibitem [{\citenamefont {Coloma}\ \emph {et~al.}(2015)\citenamefont {Coloma},
  \citenamefont {Dobrescu}, \citenamefont {Frugiuele},\ and\ \citenamefont
  {Harnik}}]{Coloma:2015pih}%
  \BibitemOpen
  \bibfield  {author} {\bibinfo {author} {\bibfnamefont {P.}~\bibnamefont
  {Coloma}}, \bibinfo {author} {\bibfnamefont {B.~A.}\ \bibnamefont
  {Dobrescu}}, \bibinfo {author} {\bibfnamefont {C.}~\bibnamefont {Frugiuele}},
  \ and\ \bibinfo {author} {\bibfnamefont {R.}~\bibnamefont {Harnik}},\
  }\href@noop {} {\  (\bibinfo {year} {2015})},\ \Eprint
  {http://arxiv.org/abs/1512.03852} {arXiv:1512.03852 [hep-ph]} \BibitemShut
  {NoStop}%
\bibitem [{\citenamefont {Pospelov}\ \emph {et~al.}(2008)\citenamefont
  {Pospelov}, \citenamefont {Ritz},\ and\ \citenamefont
  {Voloshin}}]{Pospelov:2007mp}%
  \BibitemOpen
  \bibfield  {author} {\bibinfo {author} {\bibfnamefont {M.}~\bibnamefont
  {Pospelov}}, \bibinfo {author} {\bibfnamefont {A.}~\bibnamefont {Ritz}}, \
  and\ \bibinfo {author} {\bibfnamefont {M.~B.}\ \bibnamefont {Voloshin}},\
  }\href {\doibase 10.1016/j.physletb.2008.02.052} {\bibfield  {journal}
  {\bibinfo  {journal} {Phys.Lett.}\ }\textbf {\bibinfo {volume} {B662}},\
  \bibinfo {pages} {53} (\bibinfo {year} {2008})},\ \Eprint
  {http://arxiv.org/abs/0711.4866} {arXiv:0711.4866 [hep-ph]} \BibitemShut
  {NoStop}%
\bibitem [{\citenamefont {Arkani-Hamed}\ \emph {et~al.}(2009)\citenamefont
  {Arkani-Hamed}, \citenamefont {Finkbeiner}, \citenamefont {Slatyer},\ and\
  \citenamefont {Weiner}}]{ArkaniHamed:2008qn}%
  \BibitemOpen
  \bibfield  {author} {\bibinfo {author} {\bibfnamefont {N.}~\bibnamefont
  {Arkani-Hamed}}, \bibinfo {author} {\bibfnamefont {D.~P.}\ \bibnamefont
  {Finkbeiner}}, \bibinfo {author} {\bibfnamefont {T.~R.}\ \bibnamefont
  {Slatyer}}, \ and\ \bibinfo {author} {\bibfnamefont {N.}~\bibnamefont
  {Weiner}},\ }\href {\doibase 10.1103/PhysRevD.79.015014} {\bibfield
  {journal} {\bibinfo  {journal} {Phys. Rev.}\ }\textbf {\bibinfo {volume}
  {D79}},\ \bibinfo {pages} {015014} (\bibinfo {year} {2009})},\ \Eprint
  {http://arxiv.org/abs/0810.0713} {arXiv:0810.0713 [hep-ph]} \BibitemShut
  {NoStop}%
\bibitem [{\citenamefont {Aguilar-Arevalo}\ \emph
  {et~al.}(2009{\natexlab{a}})\citenamefont {Aguilar-Arevalo} \emph
  {et~al.}}]{AguilarArevalo:2008yp}%
  \BibitemOpen
  \bibfield  {author} {\bibinfo {author} {\bibfnamefont {A.}~\bibnamefont
  {Aguilar-Arevalo}} \emph {et~al.} (\bibinfo {collaboration} {MiniBooNE
  Collaboration}),\ }\href {\doibase 10.1103/PhysRevD.79.072002} {\bibfield
  {journal} {\bibinfo  {journal} {Phys.Rev.}\ }\textbf {\bibinfo {volume}
  {D79}},\ \bibinfo {pages} {072002} (\bibinfo {year} {2009}{\natexlab{a}})},\
  \Eprint {http://arxiv.org/abs/0806.1449} {arXiv:0806.1449 [hep-ex]}
  \BibitemShut {NoStop}%
\bibitem [{\citenamefont {Aguilar-Arevalo}\ \emph
  {et~al.}(2010{\natexlab{a}})\citenamefont {Aguilar-Arevalo} \emph
  {et~al.}}]{AguilarArevalo:2010cx}%
  \BibitemOpen
  \bibfield  {author} {\bibinfo {author} {\bibfnamefont {A.~A.}\ \bibnamefont
  {Aguilar-Arevalo}} \emph {et~al.} (\bibinfo {collaboration} {MiniBooNE}),\
  }\href {\doibase 10.1103/PhysRevD.82.092005} {\bibfield  {journal} {\bibinfo
  {journal} {Phys. Rev.}\ }\textbf {\bibinfo {volume} {D82}},\ \bibinfo {pages}
  {092005} (\bibinfo {year} {2010}{\natexlab{a}})},\ \Eprint
  {http://arxiv.org/abs/1007.4730} {arXiv:1007.4730 [hep-ex]} \BibitemShut
  {NoStop}%
\bibitem [{\citenamefont {Aguilar-Arevalo}\ \emph {et~al.}(2015)\citenamefont
  {Aguilar-Arevalo} \emph {et~al.}}]{Aguilar-Arevalo:2013nkf}%
  \BibitemOpen
  \bibfield  {author} {\bibinfo {author} {\bibfnamefont {A.}~\bibnamefont
  {Aguilar-Arevalo}} \emph {et~al.} (\bibinfo {collaboration} {MiniBooNE}),\
  }\href {\doibase 10.1103/PhysRevD.91.012004} {\bibfield  {journal} {\bibinfo
  {journal} {Phys.Rev.}\ }\textbf {\bibinfo {volume} {D91}},\ \bibinfo {pages}
  {012004} (\bibinfo {year} {2015})},\ \Eprint {http://arxiv.org/abs/1309.7257}
  {arXiv:1309.7257 [hep-ex]} \BibitemShut {NoStop}%
\bibitem [{\citenamefont {Aguilar-Arevalo}\ \emph
  {et~al.}(2009{\natexlab{b}})\citenamefont {Aguilar-Arevalo} \emph
  {et~al.}}]{AguilarArevalo:2008qa}%
  \BibitemOpen
  \bibfield  {author} {\bibinfo {author} {\bibfnamefont {A.~A.}\ \bibnamefont
  {Aguilar-Arevalo}} \emph {et~al.} (\bibinfo {collaboration} {MiniBooNE}),\
  }\href {\doibase 10.1016/j.nima.2008.10.028} {\bibfield  {journal} {\bibinfo
  {journal} {Nucl. Instrum. Meth.}\ }\textbf {\bibinfo {volume} {A599}},\
  \bibinfo {pages} {28} (\bibinfo {year} {2009}{\natexlab{b}})},\ \Eprint
  {http://arxiv.org/abs/0806.4201} {arXiv:0806.4201 [hep-ex]} \BibitemShut
  {NoStop}%
\bibitem [{\citenamefont {Aguilar-Arevalo}\ \emph
  {et~al.}(2010{\natexlab{b}})\citenamefont {Aguilar-Arevalo} \emph
  {et~al.}}]{AguilarArevalo:2010zc}%
  \BibitemOpen
  \bibfield  {author} {\bibinfo {author} {\bibfnamefont {A.~A.}\ \bibnamefont
  {Aguilar-Arevalo}} \emph {et~al.} (\bibinfo {collaboration} {MiniBooNE}),\
  }\href {\doibase 10.1103/PhysRevD.81.092005} {\bibfield  {journal} {\bibinfo
  {journal} {Phys. Rev.}\ }\textbf {\bibinfo {volume} {D81}},\ \bibinfo {pages}
  {092005} (\bibinfo {year} {2010}{\natexlab{b}})},\ \Eprint
  {http://arxiv.org/abs/1002.2680} {arXiv:1002.2680 [hep-ex]} \BibitemShut
  {NoStop}%
\bibitem [{\citenamefont {Katori}\ and\ \citenamefont
  {Martini}(2016)}]{Katori:2016yel}%
  \BibitemOpen
  \bibfield  {author} {\bibinfo {author} {\bibfnamefont {T.}~\bibnamefont
  {Katori}}\ and\ \bibinfo {author} {\bibfnamefont {M.}~\bibnamefont
  {Martini}},\ }\href@noop {} {\  (\bibinfo {year} {2016})},\ \Eprint
  {http://arxiv.org/abs/1611.07770} {arXiv:1611.07770 [hep-ph]} \BibitemShut
  {NoStop}%
\bibitem [{\citenamefont {deNiverville}\ \emph {et~al.}(2017)\citenamefont
  {deNiverville}, \citenamefont {Chen}, \citenamefont {Pospelov},\ and\
  \citenamefont {Ritz}}]{deNiverville:2016rqh}%
  \BibitemOpen
  \bibfield  {author} {\bibinfo {author} {\bibfnamefont {P.}~\bibnamefont
  {deNiverville}}, \bibinfo {author} {\bibfnamefont {C.-Y.}\ \bibnamefont
  {Chen}}, \bibinfo {author} {\bibfnamefont {M.}~\bibnamefont {Pospelov}}, \
  and\ \bibinfo {author} {\bibfnamefont {A.}~\bibnamefont {Ritz}},\ }\href
  {\doibase 10.1103/PhysRevD.95.035006} {\bibfield  {journal} {\bibinfo
  {journal} {Phys. Rev.}\ }\textbf {\bibinfo {volume} {D95}},\ \bibinfo {pages}
  {035006} (\bibinfo {year} {2017})},\ \Eprint
  {http://arxiv.org/abs/1609.01770} {arXiv:1609.01770 [hep-ph]} \BibitemShut
  {NoStop}%
\bibitem [{\citenamefont {Aguilar-Arevalo}(2008)}]{AlexisThesis}%
  \BibitemOpen
  \bibfield  {author} {\bibinfo {author} {\bibfnamefont {A.}~\bibnamefont
  {Aguilar-Arevalo}},\ }\emph {\bibinfo {title} {An Improved Neutrino
  Oscillations Analysis of the MiniBooNE Data}},\ \href@noop {} {Ph.D.
  thesis},\ \bibinfo  {school} {Columbia University} (\bibinfo {year}
  {2008})\BibitemShut {NoStop}%
\bibitem [{\citenamefont {Aguilar-Arevalo}\ \emph {et~al.}(2013)\citenamefont
  {Aguilar-Arevalo} \emph {et~al.}}]{Aguilar-Arevalo:2013pmq}%
  \BibitemOpen
  \bibfield  {author} {\bibinfo {author} {\bibfnamefont {A.~A.}\ \bibnamefont
  {Aguilar-Arevalo}} \emph {et~al.} (\bibinfo {collaboration} {MiniBooNE}),\
  }\href {\doibase 10.1103/PhysRevLett.110.161801} {\bibfield  {journal}
  {\bibinfo  {journal} {Phys. Rev. Lett.}\ }\textbf {\bibinfo {volume} {110}},\
  \bibinfo {pages} {161801} (\bibinfo {year} {2013})},\ \Eprint
  {http://arxiv.org/abs/1207.4809} {arXiv:1207.4809 [hep-ex]} \BibitemShut
  {NoStop}%
\bibitem [{\citenamefont {Pospelov}(2009)}]{Pospelov:2008zw}%
  \BibitemOpen
  \bibfield  {author} {\bibinfo {author} {\bibfnamefont {M.}~\bibnamefont
  {Pospelov}},\ }\href {\doibase 10.1103/PhysRevD.80.095002} {\bibfield
  {journal} {\bibinfo  {journal} {Phys. Rev.}\ }\textbf {\bibinfo {volume}
  {D80}},\ \bibinfo {pages} {095002} (\bibinfo {year} {2009})},\ \Eprint
  {http://arxiv.org/abs/0811.1030} {arXiv:0811.1030 [hep-ph]} \BibitemShut
  {NoStop}%
\bibitem [{\citenamefont {Fayet}(2007)}]{Fayet:2007ua}%
  \BibitemOpen
  \bibfield  {author} {\bibinfo {author} {\bibfnamefont {P.}~\bibnamefont
  {Fayet}},\ }\href {\doibase 10.1103/PhysRevD.75.115017} {\bibfield  {journal}
  {\bibinfo  {journal} {Phys. Rev.}\ }\textbf {\bibinfo {volume} {D75}},\
  \bibinfo {pages} {115017} (\bibinfo {year} {2007})},\ \Eprint
  {http://arxiv.org/abs/hep-ph/0702176} {arXiv:hep-ph/0702176 [HEP-PH]}
  \BibitemShut {NoStop}%
\bibitem [{\citenamefont {Essig}\ \emph {et~al.}(2013)\citenamefont {Essig},
  \citenamefont {Mardon}, \citenamefont {Papucci}, \citenamefont {Volansky},\
  and\ \citenamefont {Zhong}}]{Essig:2013vha}%
  \BibitemOpen
  \bibfield  {author} {\bibinfo {author} {\bibfnamefont {R.}~\bibnamefont
  {Essig}}, \bibinfo {author} {\bibfnamefont {J.}~\bibnamefont {Mardon}},
  \bibinfo {author} {\bibfnamefont {M.}~\bibnamefont {Papucci}}, \bibinfo
  {author} {\bibfnamefont {T.}~\bibnamefont {Volansky}}, \ and\ \bibinfo
  {author} {\bibfnamefont {Y.-M.}\ \bibnamefont {Zhong}},\ }\href {\doibase
  10.1007/JHEP11(2013)167} {\bibfield  {journal} {\bibinfo  {journal} {JHEP}\
  }\textbf {\bibinfo {volume} {1311}},\ \bibinfo {pages} {167} (\bibinfo {year}
  {2013})},\ \Eprint {http://arxiv.org/abs/1309.5084} {arXiv:1309.5084
  [hep-ph]} \BibitemShut {NoStop}%
\bibitem [{\citenamefont {Agnese}\ \emph {et~al.}(2016)\citenamefont {Agnese}
  \emph {et~al.}}]{Agnese:2015nto}%
  \BibitemOpen
  \bibfield  {author} {\bibinfo {author} {\bibfnamefont {R.}~\bibnamefont
  {Agnese}} \emph {et~al.} (\bibinfo {collaboration} {SuperCDMS}),\ }\href
  {\doibase 10.1103/PhysRevLett.116.071301} {\bibfield  {journal} {\bibinfo
  {journal} {Phys. Rev. Lett.}\ }\textbf {\bibinfo {volume} {116}},\ \bibinfo
  {pages} {071301} (\bibinfo {year} {2016})},\ \Eprint
  {http://arxiv.org/abs/1509.02448} {arXiv:1509.02448 [astro-ph.CO]}
  \BibitemShut {NoStop}%
\bibitem [{\citenamefont {Angloher}\ \emph {et~al.}(2016)\citenamefont
  {Angloher} \emph {et~al.}}]{Angloher:2015ewa}%
  \BibitemOpen
  \bibfield  {author} {\bibinfo {author} {\bibfnamefont {G.}~\bibnamefont
  {Angloher}} \emph {et~al.} (\bibinfo {collaboration} {CRESST}),\ }\href
  {\doibase 10.1140/epjc/s10052-016-3877-3} {\bibfield  {journal} {\bibinfo
  {journal} {Eur. Phys. J.}\ }\textbf {\bibinfo {volume} {C76}},\ \bibinfo
  {pages} {25} (\bibinfo {year} {2016})},\ \Eprint
  {http://arxiv.org/abs/1509.01515} {arXiv:1509.01515 [astro-ph.CO]}
  \BibitemShut {NoStop}%
\bibitem [{\citenamefont {Banerjee}\ \emph {et~al.}(2017)\citenamefont
  {Banerjee} \emph {et~al.}}]{Banerjee:2016tad}%
  \BibitemOpen
  \bibfield  {author} {\bibinfo {author} {\bibfnamefont {D.}~\bibnamefont
  {Banerjee}} \emph {et~al.} (\bibinfo {collaboration} {NA64}),\ }\href
  {\doibase 10.1103/PhysRevLett.118.011802} {\bibfield  {journal} {\bibinfo
  {journal} {Phys. Rev. Lett.}\ }\textbf {\bibinfo {volume} {118}},\ \bibinfo
  {pages} {011802} (\bibinfo {year} {2017})},\ \Eprint
  {http://arxiv.org/abs/1610.02988} {arXiv:1610.02988 [hep-ex]} \BibitemShut
  {NoStop}%
\bibitem [{\citenamefont {Essig}\ \emph {et~al.}(2012)\citenamefont {Essig},
  \citenamefont {Manalaysay}, \citenamefont {Mardon}, \citenamefont
  {Sorensen},\ and\ \citenamefont {Volansky}}]{Essig:2012yx}%
  \BibitemOpen
  \bibfield  {author} {\bibinfo {author} {\bibfnamefont {R.}~\bibnamefont
  {Essig}}, \bibinfo {author} {\bibfnamefont {A.}~\bibnamefont {Manalaysay}},
  \bibinfo {author} {\bibfnamefont {J.}~\bibnamefont {Mardon}}, \bibinfo
  {author} {\bibfnamefont {P.}~\bibnamefont {Sorensen}}, \ and\ \bibinfo
  {author} {\bibfnamefont {T.}~\bibnamefont {Volansky}},\ }\href {\doibase
  10.1103/PhysRevLett.109.021301} {\bibfield  {journal} {\bibinfo  {journal}
  {Phys. Rev. Lett.}\ }\textbf {\bibinfo {volume} {109}},\ \bibinfo {pages}
  {021301} (\bibinfo {year} {2012})},\ \Eprint {http://arxiv.org/abs/1206.2644}
  {arXiv:1206.2644 [astro-ph.CO]} \BibitemShut {NoStop}%
\bibitem [{\citenamefont {Davoudiasl}\ and\ \citenamefont
  {Marciano}(2015)}]{Davoudiasl:2015hxa}%
  \BibitemOpen
  \bibfield  {author} {\bibinfo {author} {\bibfnamefont {H.}~\bibnamefont
  {Davoudiasl}}\ and\ \bibinfo {author} {\bibfnamefont {W.~J.}\ \bibnamefont
  {Marciano}},\ }\href {\doibase 10.1103/PhysRevD.92.035008} {\bibfield
  {journal} {\bibinfo  {journal} {Phys. Rev.}\ }\textbf {\bibinfo {volume}
  {D92}},\ \bibinfo {pages} {035008} (\bibinfo {year} {2015})},\ \Eprint
  {http://arxiv.org/abs/1502.07383} {arXiv:1502.07383 [hep-ph]} \BibitemShut
  {NoStop}%
\end{thebibliography}
